\documentclass[%
 aip,
 amsmath,amssymb,
 reprint,%
]{revtex4-1}
\DeclareUnicodeCharacter{2032}{\ensuremath{^{\prime}}}
\usepackage[T1]{fontenc}
\usepackage[utf8]{inputenc}   
\usepackage{graphicx}
\usepackage{dcolumn}
\usepackage{bm}

\usepackage[utf8]{inputenc}
\usepackage[T1]{fontenc}
\usepackage{mathptmx}
\usepackage{etoolbox}
\usepackage{anyfontsize}

\makeatletter
\def\@email#1#2{%
 \endgroup
 \patchcmd{\titleblock@produce}
  {\frontmatter@RRAPformat}
  {\frontmatter@RRAPformat{\produce@RRAP{*#1\href{mailto:#2}{#2}}}\frontmatter@RRAPformat}
  {}{}
}%
\makeatother
\begin{document}

\preprint{AIP/123-QED}

\title{A High-Throughput Search for Stable and Magnetically Robust Fe$_3$XY$_2$ Monolayers}

\author{Soheil Ershadrad}
 \affiliation{Department of Physics and Astronomy, Uppsala University, Box-516, 75120 Uppsala, Sweden}

\author{Biplab Sanyal}
\email{biplab.sanyal@physics.uu.se}
\affiliation{Department of Physics and Astronomy, Uppsala University, Box-516, 75120 Uppsala, Sweden}

\date{\today}

\begin{abstract}
We present first principles exploration of 529 Fe$_3$XY$_2$ compounds, where $X$ and $Y$ elements are selected from the $p$-block of the periodic table. Out of the entire set, 31 compounds satisfy all criteria for energetic, dynamic, mechanical, and thermal stability. Our analysis reveals several key trends: halide-containing systems exhibit the highest average magnetic moments and the highest magnetic transition temperatures, highlighting their potential for room-temperature spintronic applications. The majority of stable compounds display perpendicular magnetic anisotropy (PMA), with Fe$_3$SiTe$_2$ exhibiting the strongest PMA among all candidates. Exchange interactions are found to be governed by a dual mechanism, direct exchange between nearest-neighbor Fe atoms and indirect, $p$-orbital-mediated exchange for second-nearest neighbors and beyond. Notably, four compounds have non-centrosymmetric crystal structures and exhibit finite spiralization constants. Among them, Fe$_3$AsBr$_2$ is predicted to host N\'eel-type skyrmions even at zero external magnetic field, as confirmed by micromagnetic simulations. These findings offer a roadmap for experimental realization of novel 2D ferromagnets with enhanced functionalities.
\end{abstract}

\maketitle

\section{Introduction}
Two-dimensional (2D) magnets have emerged as promising candidates for next-generation spintronic devices, largely due to their high surface-to-volume ratio and tunable physical properties. The reduced dimensionality in these systems leads to quantum confinement effects, often resulting in electronic and magnetic behaviors that differ markedly from their bulk counterparts\cite{gibertini2019magnetic, gong2019two}. The successful synthesis of monolayer CrI\(_3\) and CrGeTe\(_3\) in 2017~\cite{huang2017layer, gong2017discovery} sparked intense interest in the exploration and development of 2D magnetic materials. However, one of the major limitations of semiconducting 2D magnets such as CrI\(_3\) and CrGeTe\(_3\) is their relatively low magnetic transition temperatures (typically in the range of 40–60 K), which severely restricts their practical applications in room-temperature technologies. In contrast, metallic 2D magnets benefit from long-range exchange interactions, which significantly enhance their magnetic transition temperatures~\cite{ershadrad2024ab}. A notable breakthrough was the synthesis of monolayer Fe\(_3\)GeTe\(_2\) in 2018~\cite{fei2018two}, which exhibits a Curie temperature of approximately 250~K. This discovery marked a pivotal step toward the realization of practical, room-temperature applications of 2D magnetic materials in devices. Subsequent studies revealed that other members of the Fe\(_n\)GeTe\(_2\) family with higher Fe concentrations, such as Fe\(_4\)GeTe\(_2\) and Fe\(_5\)GeTe\(_2\), exhibit even higher Curie temperatures~\cite{seo2020nearly, may2019ferromagnetism, ershadrad2022unusual}. Furthermore, doping with 3$d$ transition metals has been shown to further enhance the magnetic transition temperature, offering additional tunability for spintronic applications~\cite{tian2020tunable, ghosh2024structural}. To date, several functional spintronic devices, such as spin valves and spin–orbit torque devices, have been successfully fabricated using Fe\(_n\)GeTe\(_2\) thin films~\cite{zhao2023room, ngaloy2024strong, zhao2025coexisting}. 

The recent synthesis of Fe\(_3\)GaTe\(_2\) monolayers~\cite{zhang2022above}, exhibiting a magnetic transition temperature of 300~K, suggests that elemental substitution within the Fe\(_3\)XY\(_2\) framework can be an effective strategy for designing superior 2D magnets. 

In addition to limitations in magnetic transition temperatures, both Fe\(_3\)GeTe\(_2\) and Fe\(_3\)GaTe\(_2\) are known to be highly air-sensitive, with surface oxidation significantly degrading their magnetic properties~\cite{gweon2021exchange}. Furthermore, based on the periodic table of elemental scarcity~\cite{EuChemS2019}, tellurium availability is projected to face serious depletion within the next 100 years. Therefore, the prediction and synthesis of Te-free two-dimensional magnets are crucial for ensuring long-term viability in industrial applications.

High-throughput (HTP) screening has emerged as a powerful strategy for systematically identifying promising candidates in materials discovery. While several large-scale materials databases exist, they primarily focus on bulk (3D) compounds~\cite{zagorac2019recent, jain2013commentary}. Consequently, most existing HTP studies rely on data mining from these databases, which limits the systematic exploration of 2D magnets to few compounds that are already known or previously reported~\cite{haastrup2018computational, mounet2018two}. We address this gap by developing a custom HTP platform that enables the autonomous generation, structural optimization, and magnetic property modeling of a wide range of compounds. Similar HTP methodologies have proven successful in other material classes, such as MAX phases and MXenes~\cite{carlsson2024systematic, helmer2024computational, bjork2024two}. By applying a similar framework to 2D magnets, we not only uncover trends and hidden correlations that are difficult to detect from studies of individual compounds, but also generate a rich dataset that can support future data-driven approaches, including machine learning-based materials prediction.

In this study, we employ HTP first-principles calculations to explore 529 possible Fe\(_3\)XY\(_2\) compounds in search of robust and stable two-dimensional magnetic materials. Our goal is to identify promising alternatives to Fe\(_3\)GeTe\(_2\) and Fe\(_3\)GaTe\(_2\) that not only overcome limitations associated with these compounds, but also exhibit theoretically superior magnetic properties. The results of this work can serve as a roadmap for experimental efforts, offering guidance on which Fe\(_3\)XY\(_2\) compounds are more likely to be synthesizable and possess the targeted magnetic and structural properties.

\section{Results and Discussion}

 \begin{figure*}[htp!]
 \centering
 \includegraphics[width=\linewidth]{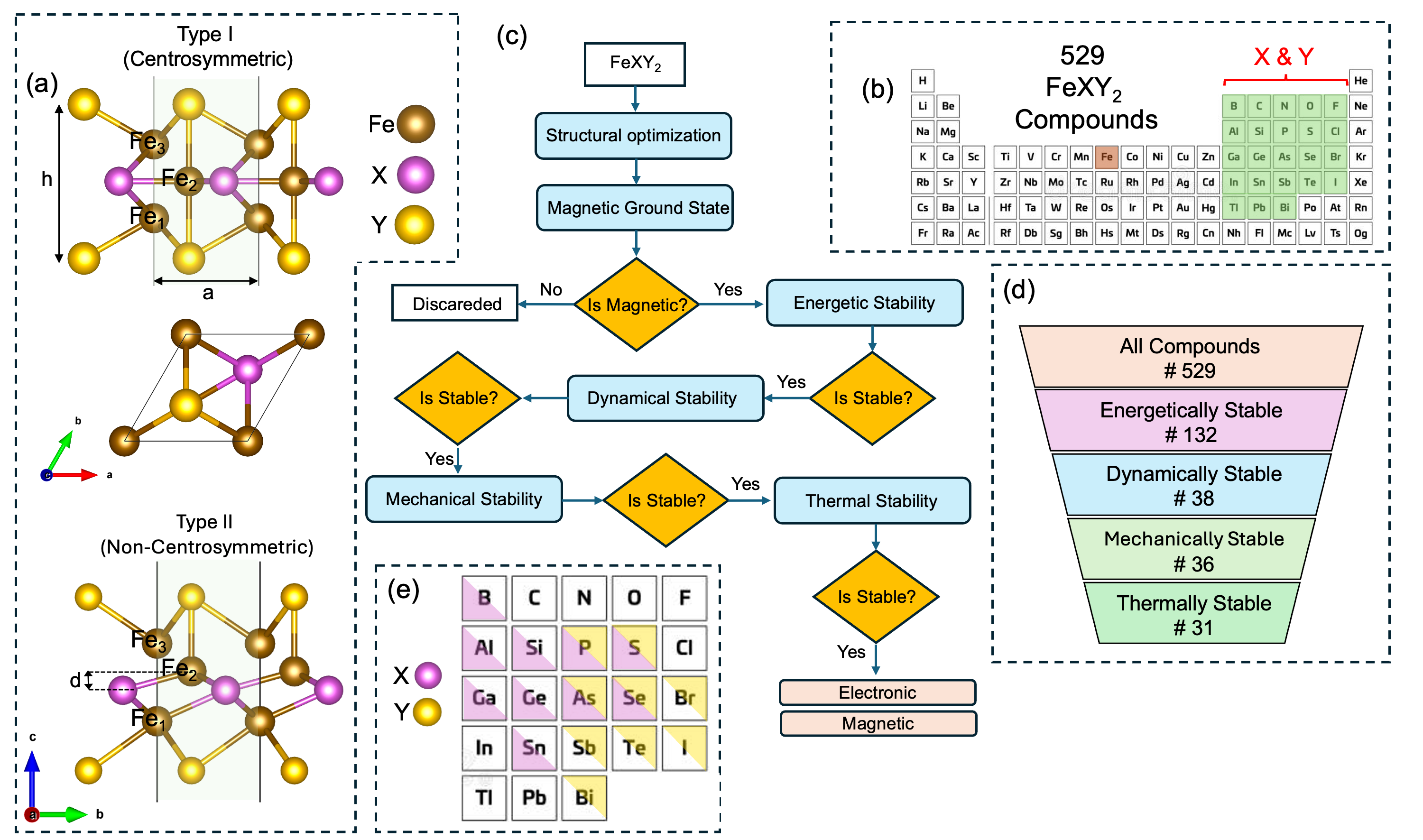}
  \caption{(a) Structural configuration of the Fe$_{3}$XY$_{2}$ monolayers depicted from both top and side views. Type I and Type II structures correspond to centrosymmetric and non-centrosymmetric ground states, respectively. The light-green shaded region highlights the unit cell, where $a$ denote the lattice parameter, $h$ represents the thickness. (b) Periodic table highlighting in green the elements that can occupy the X and Y sites, yielding 529 possible compounds. (c) Flowchart of high-throughput identification of stable compounds.  (d) Schematic showing the number of stable compounds remaining after successive stability criteria checks, yielding 31 stable compounds. (e) $p$-block of the periodic table, with X-site elements of stable compounds highlighted in pink and Y-site elements of stable compounds highlighted in gold.
}
  \label{fig1}
\end{figure*} 

For HTP exploration of Fe$_3$XY$_2$ compounds, a structural symmetry analogous to that of the experimentally synthesized Fe$_3$GeTe$_2$ monolayer was adopted as the initial reference. After structural optimizations, two distinct structural types were identified across the studied compounds. As shown schematically in Fig.~\ref{fig1}(a), the centrosymmetric Type~I structure have space group \textit{P$\bar{6}$m2} (No.~187), whereas the non-centrosymmetric Type~II structure belongs to space group \textit{P3m1} (No.~156). In Type~I compounds, similar to Fe$_3$GeTe$_2$ and Fe$_3$GaTe$_2$, the Fe$_2$ sublattice and the $X$ atoms lie within the same horizontal plane. In contrast, Type~II compounds exhibit a vertical displacement between the Fe$_2$ and $X$ atoms, denoted by \( d \), indicating a break in mirror symmetry along the out-of-plane direction. 
Fig.~\ref{fig1}(b) presents the periodic table, highlighting in green the $p$-block elements considered for the $X$ and $Y$ sites in the Fe$_3$XY$_2$ compounds. Based on all possible combinations, a total of 529 unique Fe$_3$XY$_2$ compositions were generated and systematically investigated. Fig.~\ref{fig1}(c) illustrates the flowchart outlining the high-throughput screening procedure used to identify stable compounds. Following structural optimization, the magnetic ground state of each Fe$_3$XY$_2$ composition was determined. While several non-magnetic and ferrimagnetic configurations were identified, none were found to be stable. All compounds that passed the stability criteria exhibit a ferromagnetic (FM) ground state. Fig.~\ref{fig1}(d) summarizes the screening results. Out of the initial 529 Fe$_3$XY$_2$ compounds, 132 were found to be energetically stable based on cohesive and formation energy analysis. Among these, 38 passed the dynamical stability check via phonon calculations. Two compounds failed to meet the mechanical stability criteria. Subsequent \textit{ab initio} molecular dynamics (AIMD) simulations were performed to assess thermal stability, ultimately identifying 31 compounds that satisfy all four stability criteria. 
A summary of the structural and magnetic properties of the 31 stable Fe$_3$XY$_2$ compounds, along with their chemical formulas, is provided in Table~\ref{tab1}. Fig.~\ref{fig1}(e) highlights the elements occupying the $X$ and $Y$ sites in the stable compounds, marked in pink and gold, respectively. It can be noted that the $X$ site is never occupied by halogen elements, whereas the $Y$ site is typically occupied by heavier elements from the pnictogen, chalcogen, and halogen groups.
Details of the stability assessments, including phonon spectra (Fig. S1) and AIMD energy profiles (Fig. S2), are provided in the Supplemental Materials. The cohesive and formation energies are summarized in Table~\ref{tab1}. A positive cohesive energy indicates that the atoms prefer to bind together, while a negative formation energy implies that the Fe$_3$XY$_2$ compounds are energetically more favorable than their constituent bulk counterparts in their elemental form. The elastic constants \( C_{11} \), \( C_{12} \), and \( C_{66} \) are also presented in Table~\ref{tab1}. All listed compounds satisfy the Born–Huang \cite{born1954dynamical,mouhat2014necessary} mechanical stability criteria for hexagonal monolayers (i.e. $C_{11} > \mid{C_{12}}\mid$ and $C_{66} > 0$)~\cite{varjovi2023two}. 

\begin{table*}[t!]
\centering
\fontsize{6.0}{6.5}\selectfont
\setlength{\tabcolsep}{3pt}
\renewcommand{\arraystretch}{1.5}
\caption{The optimized lattice constant $a$ (\AA); thickness $c$ (\AA); symmetry (Cen.: centrosymmetric, Non. : non-centrosymmetric); 
pair distances $d_{13}$ (Fe$_{1}$--Fe$_{3}$, \AA) and $d_{12}$ (Fe$_{1}$--Fe$_{2}$, \AA); cohesive energy $E_{\text{coh}}$ (eV/atom); 
formation energy $E_{f}$ (meV/atom); elastic constants $C_{ij}$ (GPa); spin magnetic moment per Fe atom 
$M^{\text{spin}}_{\text{Fe}}$ ($\mu_B$); orbital magnetic moment per Fe atom $M^{\text{orb}}_{\text{Fe}}$ ($\mu_B$); 
exchange parameters $J_{ij}$ (meV); exchange stiffness $A$ (meV.\AA$^{2}$); spiralization constant $D$ (meV.\AA); 
magnetic anisotropy energy (MAE, meV/f.u.); and Curie temperature $T_c$ (K) for \textit{Fe}$_3$\textit{X}\textit{Y}$_2$ monolayers.
}
\begin{tabular}{l|c|c|c|c|c|c|c|c|c|c|c|c|c|c|c|c|c|c|c|c|c}
\hline\hline 
Struct. & $a$ & $c$ & Sym. & $d_{13}$ & $d_{12}$ & $E_{coh}$ & $E_{f}$ & $C_{11}$ & $C_{12}$ & $C_{66}$ & M$^{\text{spin}}_{\text{Fe$_{1}$}}$ & M$^{\text{spin}}_{\text{Fe$_{2}$}}$ & M$^{\text{orb}}_{\text{Fe$_{1}$}}$ & M$^{\text{orb}}_{\text{Fe$_{2}$}}$ & $J_{13}$ & $J_{12}$ & $J_{22}$ & $A$ & $D$ & MAE & $T_c$ \\ 
        & (\AA) & (\AA) & -- & (\AA) & (\AA) & (eV/atom) & (eV/atom) & (GPa) & (GPa) & (GPa) & ($\mu_B$) & ($\mu_B$) & ($\mu_B$) & ($\mu_B$) & (meV) & (meV) & (meV) & (meV.\AA$^{2}$) & (meV.\AA) & (meV/f.u.) & (K) \\
\hline
Fe$_3$P$_3$   & 3.71 & 4.61 & Cen. & 2.80 & 2.56 & 5.28 & -0.16 & 741.49 & 578.76 & 81.37 & 2.21 & 0.36 & 0.034 & -0.007& 58.21 & 6.17  & -2.10  & 292.28 & 0.00 & 0.26 & 247 \\
Fe$_3$PAs$_2$  & 3.77 & 4.88 & Cen. & 2.64 & 2.55 & 5.04 & -0.15 & 653.54 & 537.13 & 58.21 & 2.34 & 0.60 & 0.031 & 0.002 & 59.21 & 11.89 & -1.86  & 558.91 & 0.00 & 0.04 & 671 \\
Fe$_3$PSb$_2$   & 3.84 & 5.32 & Cen. & 2.49 & 2.55 & 4.81 & -0.09 & 623.39 & 525.81 & 48.79 & 2.32 & 0.92 & 0.038 & 0.001 & 64.07 & 16.51 & -3.74  & 531.87 & 0.00 & -0.87 & 579 \\
Fe$_3$PBi$_2$   & 3.87 & 5.60 & Cen. & 2.44 & 2.54 & 4.65 & -0.04 & 527.35 & 455.40 & 35.98 & 2.41 & 1.26 & 0.071 & -0.004& 67.38 & 22.26 & -5.89  & 565.85 & 0.00 & 1.23 & 610 \\
Fe$_3$BS$_2$    & 3.59 & 4.48 & Cen. & 2.52 & 2.43 & 5.17 & -0.23 & 658.06 & 469.16 & 94.45 & 2.37 & 0.92 & 0.020 & 0.013 & 55.15 & 9.58  & -2.89  & 432.52 &  0.00 & -0.34 & 606 \\
Fe$_3$GaS$_2$  & 3.95 & 4.38 & Cen. & 2.57 & 3.95 & 4.51 & -0.20 & 636.94 & 309.01 & 163.96 & 2.75 & 0.45 & 0.028 & 0.012 & 56.10 & 7.36  & -0.98  & 473.82 & 0.00 & -0.37 & 605 \\
Fe$_3$SiS$_2$  & 3.85 & 4.47 & Cen. & 2.60 & 2.57 & 5.00 & -0.27 & 727.60 & 446.59 & 140.51 & 2.74 & 0.89 & 0.027 & 0.011 & 91.25 & 15.99 & -8.11  & 314.65 &  0.00 & -0.08 & 255 \\
Fe$_3$GeS$_2$  & 3.93 & 4.48 & Cen. & 2.66 & 2.63 & 4.74 & -0.17 & 696.00 & 409.75 & 143.13 & 2.85 & 0.89 & 0.027 & 0.015 & 99.19 & 15.87 & -4.06  & 446.98 & 0.00 & -0.44 & 410 \\
Fe$_3$AsS$_2$  & 3.93 & 4.56 & Non. & 2.76 & 2.62 & 4.74 & -0.14 & 574.01 & 420.05 & 76.72 & 2.87 & 1.14 & 0.030 & 0.015 & 75.48 & 25.76 & -11.20 & 376.98 & 0.59 & -0.07 & 301 \\
Fe$_3$BSe$_2$  & 3.65 & 4.81 & Cen. & 2.43 & 2.43 & 4.90 & -0.17 & 581.49 & 435.22 & 73.13 & 2.39 & 1.19 & 0.025 & 0.011 & 71.43 & 21.97 & -2.60  & 516.21 & 0.00 & 0.33 & 721 \\
Fe$_3$AlSe$_2$  & 4.00 & 4.65 & Cen. & 2.46 & 2.62 & 4.46 & -0.23 & 631.09 & 290.53 & 170.28 & 2.64 & 0.91 & 0.034 & 0.014 & 57.11 & 10.89 & -0.73  & 519.55 & 0.00 & 0.27 & 629 \\
Fe$_3$GaSe$_2$  & 4.01 & 4.67 & Cen. & 2.48 & 2.62 & 4.31 & -0.21 & 637.31 & 294.73 & 171.29 & 2.68 & 0.85 & 0.036 & 0.014 & 52.32 & 10.37 & -1.58  & 482.27 & 0.00 & 0.83 & 595 \\
Fe$_3$SiSe$_2$  & 3.91 & 4.76 & Cen. & 2.50 & 2.58 & 4.79 & -0.28 & 681.80 & 430.39 & 125.71 & 2.68 & 1.08 & 0.030 & 0.010 & 90.82 & 21.49 & -8.15  & 410.02 & 0.00 & 0.93 & 446 \\
Fe$_3$GeSe$_2$  & 3.99 & 4.77 & Cen. & 2.55 & 2.64 & 4.54 & -0.19 & 644.09 & 391.26 & 126.42 & 2.77 & 1.13 & 0.034 & 0.016 & 100.66& 22.99 & -4.88  & 530.27 & 0.00 & 0.47 & 551 \\
Fe$_3$SnSe$_2$  & 4.20 & 4.75 & Cen. & 2.60 & 2.75 & 4.27 & -0.02 & 549.05 & 360.01 & 94.52 & 2.85 & 1.32 & 0.036 & 0.032 & 94.03 & 20.39 & -5.66  & 553.24 & 0.00 & -0.55 & 543 \\
Fe$_3$BTe$_2$   & 3.69 & 5.34 & Cen. & 2.38 & 2.44 & 4.65 & -0.04 & 329.63 & 139.34 & 95.14 & 2.29 & 1.64 & 0.033 & 0.011 & 100.62& 28.01 & -4.68  & 372.35 & 0.00 & 3.28 & 457 \\
Fe$_3$AlTe$_2$ & 4.07 & 5.04 & Cen. & 2.40 & 2.64 & 4.29 & -0.18 & 625.56 & 288.70 & 168.43 & 2.52 & 1.42 & 0.044 & 0.017 & 77.16 & 22.01 & -0.58  & 630.81 & 0.00 & 0.83 & 772 \\
Fe$_3$GaTe$_2$ & 4.07 & 5.06 & Cen. & 2.42 & 2.64 & 4.14 & -0.16 & 626.39 & 277.36 & 174.51 & 2.54 & 1.41 & 0.044 & 0.022 & 85.10 & 22.32 & -1.05  & 609.97 & 0.00 & 0.55 & 667 \\
Fe$_3$SiTe$_2$ & 3.97 & 5.14 & Cen. & 2.43 & 2.59 & 4.61 & -0.21 & 633.35 & 377.17 & 128.09 & 2.52 & 1.36 & 0.037 & 0.013 & 93.53 & 31.65 & -11.49 & 395.85 & 0.00 & 5.03 & 453 \\
Fe$_3$GeTe$_2$ & 4.05 & 5.14 & Cen. & 2.47 & 2.65 & 4.37 & -0.13 & 609.45 & 356.01 & 126.72 & 2.63 & 1.46 & 0.042 & 0.022 & 87.82 & 30.43 & -5.64  & 545.60 & 0.00 & 4.15 & 611 \\
Fe$_3$SnTe$_2$ & 4.26 & 5.10 & Cen. & 2.52 & 2.76 & 4.13 & 0.00  & 547.25 & 321.94 & 112.66 & 2.73 & 1.47 & 0.045 & 0.036 & 96.46 & 27.79 & -4.42  & 616.69 & 0.00 & 1.63 & 619 \\
Fe$_3$PTe$_2$  & 3.76 & 5.55 & Non. & 2.45 & 2.45 & 4.61 & -0.22 & 588.83 & 329.14 & 129.85 & 2.56 & 1.31 & 0.038 & 0.008 & 43.92 & 43.92 & -6.30  & 174.58 & 8.58 & 1.21 & 272 \\
Fe$_3$AlBr$_2$ & 4.07 & 5.22 & Cen. & 2.33 & 2.62 & 3.75 & -0.14 & 476.35 & 208.76 & 133.80 & 2.79 & 2.06 & 0.036 & 0.027 & 86.68 & 44.04 & -2.44  & 794.61 & 0.00 & -0.33 & 906 \\
Fe$_3$GaBr$_2$ & 4.08 & 5.24 & Cen. & 2.34 & 2.63 & 3.63 & -0.15 & 462.42 & 199.73 & 131.34 & 2.87 & 2.13 & 0.037 & 0.036 & 95.05 & 40.97 & -3.91  & 763.88 & 0.00 & -0.28 & 852 \\
Fe$_3$AsBr$_2$ & 3.92 & 5.72 & Non. & 2.44 & 2.44 & 3.88 & -0.12 & 439.32 & 257.12 & 91.10 & 2.70 & 1.61 & 0.048 & 0.018 & 53.81 & 53.74 & -16.76 & 327.54 & 9.31 & 0.53 & 447 \\
Fe$_3$SBr$_2$  & 3.99 & 5.46 & Cen. & 2.51 & 2.62 & 3.89 & -0.21 & 333.81 & 257.41 & 38.20 & 3.24 & 2.53 & 0.031 & 0.019 & 77.56 & 51.04 & -10.15 & 334.88 & 0.00 & -0.44 & 231 \\
Fe$_3$AlI$_2$ & 4.12 & 5.48 & Cen. & 2.33 & 2.65 & 3.66 & -0.10 & 520.62 & 215.36 & 152.63 & 2.73 & 2.09 & 0.036 & 0.032 & 94.77 & 45.31 & -2.16  & 809.70 & 0.00 & -0.77 & 891 \\
Fe$_3$GaI$_2$  & 4.12 & 5.49 & Cen. & 2.34 & 2.65 & 3.52 & -0.10 & 499.19 & 183.75 & 157.72 & 2.82 & 2.09 & 0.036 & 0.040 & 110.56& 43.49 & -3.30  & 826.10 & 0.00 & 1.16 & 914 \\
Fe$_3$SiI$_2$  & 3.99 & 5.74 & Cen. & 2.37 & 2.59 & 3.95 & -0.11 & 526.16 & 236.53 & 144.81 & 2.66 & 2.04 & 0.042 & 0.031 & 61.26 & 36.51 & -15.70 & 522.02 & 0.00 & 0.74 & 605 \\
Fe$_3$GeI$_2$ & 4.10 & 5.67 & Cen. & 2.38 & 2.65 & 3.73 & -0.05 & 446.80 & 182.93 & 131.93 & 2.72 & 2.12 & 0.049 & 0.031 & 76.67 & 34.47 & -4.30  & 774.62 & 0.00 & 1.08 & 888 \\
Fe$_3$SI$_2$  & 4.06 & 5.79 & Cen. & 2.48 & 2.65 & 3.71 & -0.09 & 325.92 & 270.46 & 27.73 & 3.18 & 2.58 & 0.031 & 0.022 & 104.41& 51.52 & -8.83  & 416.89 & 0.00 & 0.02 & 165 \\
Fe$_3$SeI$_2$  & 4.20 & 5.77 & Cen. & 2.50 & 2.72 & 3.54 & -0.02 & 349.74 & 249.66 & 50.04 & 3.18 & 2.64 & 0.034 & 0.029 & 84.33 & 48.34 & -7.52  & 387.53 & 0.00 & 0.43 & 160 \\
Fe$_3$AsI$_2$  & 3.99 & 5.96 & Non. & 2.48 & 2.48 & 3.74 & -0.03 & 452.00 & 232.96 & 109.52 & 2.53 & 1.74 & 0.035 & 0.025 & 56.87 & 56.68 & -13.03 & 181.55 & 4.71 & -0.66 & 307 \\
\hline\hline
\end{tabular}
\label{tab1}
\end{table*}

 \begin{figure*}[htp!]
 \centering
 \includegraphics[width=\linewidth]{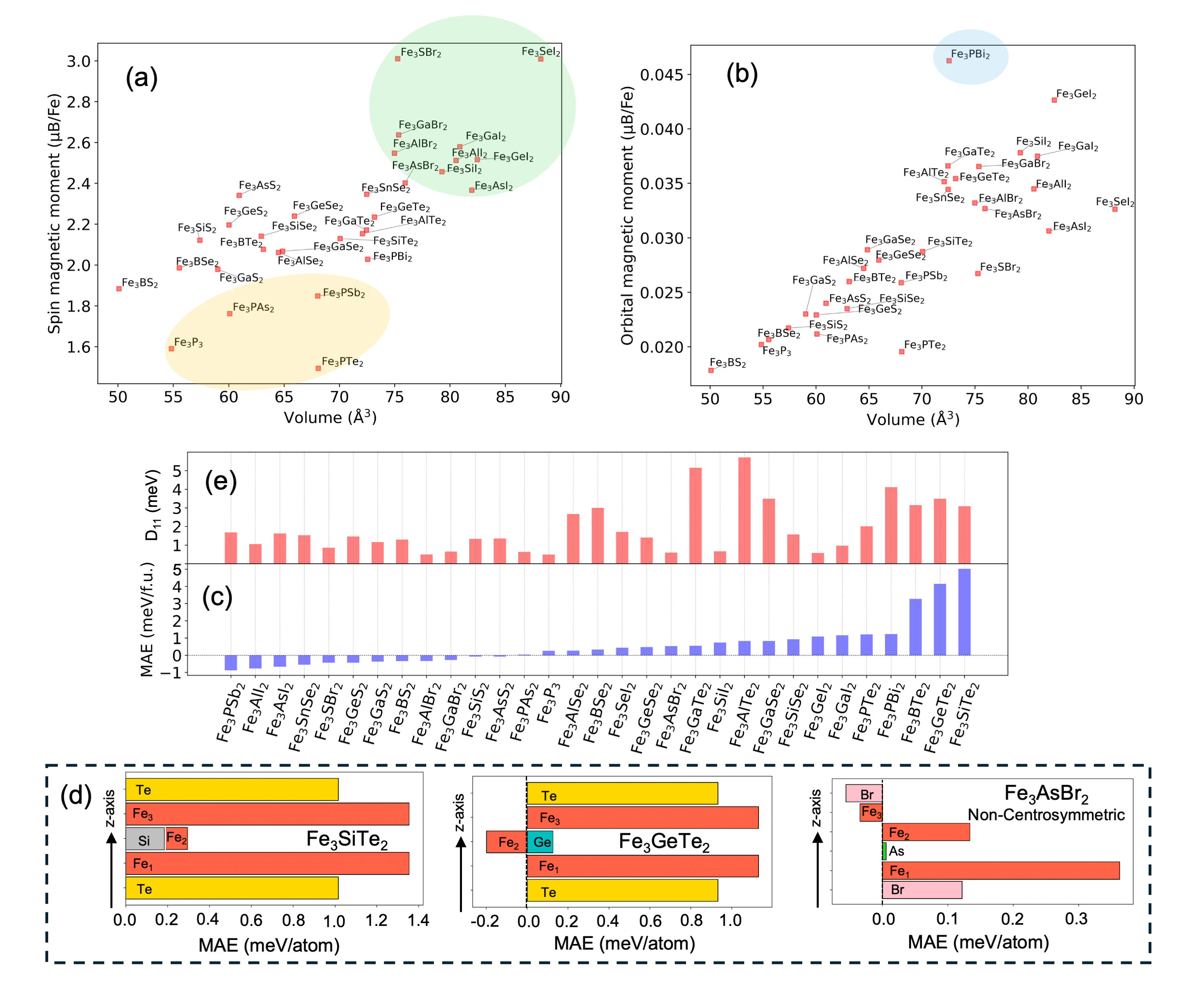}
  \caption{(a) Scatter plot of the averaged spin magnetic moment per Fe vs. cell volume, where the green zone highlights halides with the highest averaged moments and the yellow zone highlights compounds containing P at the X site with the lowest magnetic moments. (b) Scatter plot of the averaged orbital magnetic moment per Fe vs. cell volume, where the highest orbital moment is highlighted in blue. (c) Magnetic anisotropy energy (MAE) sorted in ascending order for 31 stable compounds. (d) Corresponding $D_{11}$ values. (e) Atom-resolved MAE for Fe$_3$SiTe$_2$ (the compound with the strongest MAE), Fe$_3$GeTe$_2$ (the synthesized monolayer) and Fe$_3$AsBr$_2$ (a non-centrosymmetric compound).}
  \label{fig2}
\end{figure*} 


Extracted spin and orbital moments for Fe$_{1}$ and Fe$_{2}$ sublattices are also listed in Table~\ref{tab1} (Fe$_{3}$ sublattice is symmetrically equivalent to Fe$_{1}$ in centrocymetric systems). Fig.~\ref{fig2}(a) and (b) show scatter plots of the spin and orbital magnetic moments vs. the unit cell volume, respectively. In general, both spin and orbital moments tend to increase with increasing cell volume. In general, Fe atoms tend to exhibit higher magnetic moments as the atomic volume per Fe increases, even in elemental bulk forms. This behavior is primarily attributed to the reduced bandwidth resulting from increased interatomic spacing, which enhances the localization of $d$-electrons~\cite{soulairol2010structure}. The green-shaded region in Fig.~\ref{fig2}(a) highlights the compounds exhibiting the highest average spin magnetic moments per Fe atom. Notably, all these compounds are halides, with Fe$_3$SeI$_2$ and Fe$_3$SBr$_2$ showing average spin moments of approximately 3.0~$\mu_\mathrm{B}$, significantly higher than that of Fe$_3$GeTe$_2$ (2.23~$\mu_\mathrm{B}$). A mixed valence state between $+2$ and $+3$ is expected for Fe$_3$GeTe$_2$~\cite{backes2024valence}. In this range, Fe atoms possess three to four unpaired electrons, corresponding to a spin-only magnetic moment of approximately $3.9$–$4.9~\mu_{\mathrm{B}}$, as given by the relation \(\mu_{\text{spin-only}} = \sqrt{n(n+2)}~\mu_{\mathrm{B}}\). Element-specific X-ray magnetic circular dichroism (XMCD) measurements by K. Yamagami \textit{et al.}~\cite{yamagami2021itinerant, yamagami2022enhanced} revealed that the reduced magnetic moment arises from strong hybridization between Fe~3\textit{d} orbitals and the \textit{p} orbitals of neighboring Te and Ge atoms. This hybridization leads to significant band broadening and delocalization of Fe~\textit{d}-electrons. Our recent study on 2D metallic magnets FeXZ\(_2\) (X = Nb, Ta; Z = S, Se, Te) reveals that hybridization between Fe \(d\)-orbitals and chalcogen \(p\)-orbitals plays a key role in governing the magnetic properties~\cite{ershadrad2025complex}. Accordingly, the hybridization between Fe and halogen atoms in halide-containing compounds is expected to be weaker compared to chalcogen counterparts. Considering that Br has a smaller atomic radius than Te, the larger unit cell volume observed in halides such as Fe$_3$GaBr$_2$ compared to Fe$_3$GaTe$_2$ can be attributed to the weaker hybridization between Fe and Br atoms. This reduced hybridization results in less bonding contraction, thereby allowing for a more expanded lattice. Moreover, it can be observed that compounds with P occupying the X site exhibit the smallest average magnetic moments (see the yellow highlighted zone). This trend can be attributed to the significant hybridization between the $p$-orbitals of P atoms and the $d$-orbitals of Fe atoms, which leads to a reduction in the localized magnetic moment on Fe. The orbital moment scatter plot in Fig.~\ref{fig2}(b) reveals that the orbital moments depend not only on structural parameters but also on the spin--orbit coupling (SOC) strength of the \( X \) and \( Y \) elements. Among the studied compounds, Fe\(_3\)PBi\(_2\) exhibits the highest average orbital moment per Fe atom, whereas systems containing lighter elements, such as Fe\(_3\)BS\(_2\), show a low orbital moment. 

\begin{figure*}[t!]
 \centering
 \includegraphics[width=\linewidth]{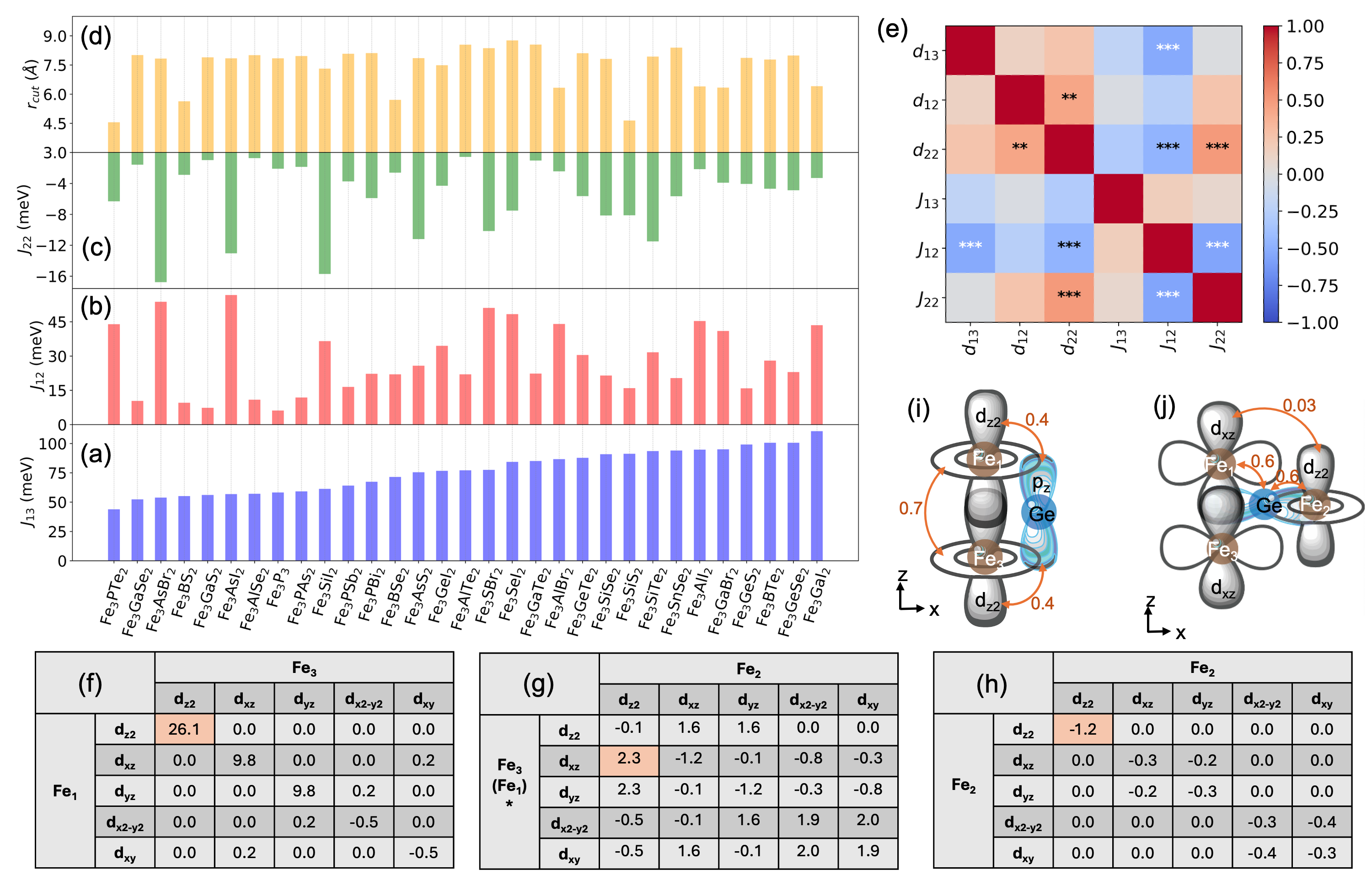}
  \caption{Analysis of Heisenberg exchange interactions. (a) J$_{13}$, sorted in ascending order for 31 stable compounds. (b) J$_{12}$ (c) J$_{22}$, negative sign indicates AFM nature of interaction, (d) the cut off radius where strength of exchange interactions drops below 1 meV. (e) Heatmap of Pearson correlation coefficients between pair distances and exchange interactions. Each box displays the correlation coefficient, with values close to \(+1\) or \(-1\) indicating strong direct or inverse relationships, respectively. Asterisks denote the statistical significance of each correlation: \(*\) for \(p < 0.1\), \(**\) for \(p < 0.05\), and \(***\) for \(p < 0.01\). (f--h) Orbital-decomposed exchange interactions for J$_{13}$ (f), J$_{12}$ (g), and J$_{22}$ (h), with the strongest interaction highlighted in orange. Positive and negative values correspond to FM and AFM interactions, respectively. (i) and (j) Schematic illustration of the $p$- and $d$-orbitals involved in the strongest interactions for J$_{13}$ and J$_{12}$, respectively. The orange double-headed arrows and the number next to them represent the Wannierization hopping parameters between orbitals, given in meV.
  }
  \label{fig3}
\end{figure*} 

Fig.~\ref{fig2}(c) presents the magnetic anisotropy energy (MAE) of the stable compounds in ascending order. A negative MAE indicates easy-plane magnetism, whereas a positive value corresponds to perpendicular (out-of-plane) easy-axis magnetism. The calculated MAE values for Fe\(_3\)GeTe\(_2\) and Fe\(_3\)GaTe\(_2\) are in excellent agreement with previously reported results~\cite{ghosh2023unraveling, marfoua2024large}. Notably, Fe\(_3\)SiTe\(_2\) exhibits a larger perpendicular MAE compared to Fe\(_3\)GeTe\(_2\), suggesting enhanced magnetic anisotropy in this compound. While the strong SOC of Te atoms is an important contributor, the enhanced MAE observed in the presence of Si suggests that magnetic anisotropy is further influenced by electronic factors, particularly \( p \)--\( d \) orbital hybridization.

Fig.~\ref{fig2}(d) shows the atom-resolved MAE for three representative cases: Fe\(_3\)SiTe\(_2\), which exhibits the strongest MAE among the studied compounds, synthesized Fe\(_3\)GeTe\(_2\), and the non-centrosymmetric system Fe\(_3\)AsBr\(_2\). Atom-resolved MAE data for all other compounds are provided in the supplementary materials (see \texttt{atom-resolved-mae.txt} in each corresponding folder). It can be observed that, in general, Fe\(_3\)SiTe\(_2\) and Fe\(_3\)GeTe\(_2\) exhibit similar atom-resolved MAE profiles, except for the direction of MAE in the Fe\(_2\) sublattice. This difference can be attributed to the variation in hybridization between the Fe\(_2\) atoms and the \( X \)-site elements. Specifically, the hybridization between Fe\(_2\) and Si promotes perpendicular anisotropy, whereas the interaction between Fe\(_2\) and Ge favors in-plane MAE. This distinction explains the overall enhancement of perpendicular anisotropy in Fe\(_3\)SiTe\(_2\). Moreover, in the non-centrosymmetric Fe\(_3\)AsBr\(_2\), an asymmetric distribution of atom-resolved MAE is observed: the Fe\(_1\), Fe\(_2\), and Br\(_\mathrm{Dn}\) sublayers exhibit positive MAE values, whereas Fe\(_3\) and Br\(_\mathrm{Up}\) show negative contributions. The competition among these opposing anisotropies leads to an overall reduction in the total MAE of the compound. In later sections, it will be shown that this reduction in MAE plays a crucial role in stabilizing skyrmions in this compound. Fig.~\ref{fig2}(e) presents the strength of the \( D_{11} \) (Dzyaloshinskii–Moriya interaction, DMI) across all dynamically stable compounds. Te-based systems generally exhibit strong \( D_{11} \) values, with Fe\(_3\)AlTe\(_2\) showing the highest at 5.71~meV. In contrast, halide compounds show relatively weak DMI; for example, Fe\(_3\)GaI\(_2\) has \( D_{11} = 0.97 \)~meV. It can be inferred that, similar to MAE, the strength of the DMI is not solely determined by the SOC strength of the constituent elements. Instead, the hybridization between atomic orbitals also plays a significant role in governing the magnitude of DMI. It should be mentioned that in centrosymmetric systems, the DMI vectors cancel due to symmetry, regardless of their individual magnitudes, and therefore chiral spin textures are not expected in these compounds.

The Heisenberg exchange interactions ($J_{ij}$) were extracted for all Fe$_3$XY$_2$ compounds, and a summary of the results is presented in Fig.~\ref{fig3}(a--d). The complete datasets of $J_{ij}$ and $D_{ij}$ as functions of interatomic distance are available in the Supplemental Materials, in corresponding folder for each compound. Fig.~\ref{fig3}(a) displays the strength of the $J_{13}$ exchange interaction, sorted in ascending order for all Fe$_3$XY$_2$ compounds. In general, $J_{13}$ is found to be strong, reaching up to 110~meV in Fe$_3$GaI$_2$. This strong interaction is recognized as the primary contributor to the elevated $T_c$ in Fe$_3$GeTe$_2$~\cite{ghosh2023unraveling}. Fig.~\ref{fig3}(b) and (c) show the values of $J_{12}$ and $J_{22}$ interactions, respectively. While both $J_{13}$ and $J_{12}$ consistently exhibit FM character, $J_{22}$ is AFM in all compounds. This behavior aligns with prior findings that intra-sublayer interactions are inherently AFM in Fe$_3$GeTe$_2$~\cite{ghosh2023unraveling}.

Although $J_{12}$ and $J_{22}$ often follow similar trends (e.g., both are strong in Fe$_3$AsBr$_2$ and Fe$_3$AsI$_2$), their behavior does not correlate with that of $J_{13}$. To further investigate potential correlations, a heatmap of Pearson correlation coefficients between exchange interactions and interatomic distances is presented in Fig.~\ref{fig3}(e). Coefficients near \(+1\) or \(-1\) indicate strong direct or inverse correlations, respectively. Statistical significance is denoted by asterisks: $^*$ for $p < 0.1$, $^{**}$ for $p < 0.05$, and $^{***}$ for $p < 0.01$.

The heatmap reveals a strong correlation between $J_{12}$ and $J_{22}$, with high statistical significance. The inverse correlation arises from the fact that $J_{22}$ is consistently antiferromagnetic (AFM) and thus carries a negative sign. In contrast, $J_{13}$ shows no meaningful correlation with either $J_{12}$ or $J_{22}$, nor with any of the pair distances, including $d_{13}$. Interestingly, $J_{12}$ exhibits a significant inverse correlation with $d_{13}$, while both $J_{12}$ and $J_{22}$ show strong correlations with $d_{22}$. These results suggest that while $J_{12}$ and $J_{22}$ are governed by geometric factors such as atomic separations, the dominant $J_{13}$ interaction follows a different, less distance-dependent mechanism. 

\begin{figure*}[t!]
 \centering
 \includegraphics[width=\linewidth]{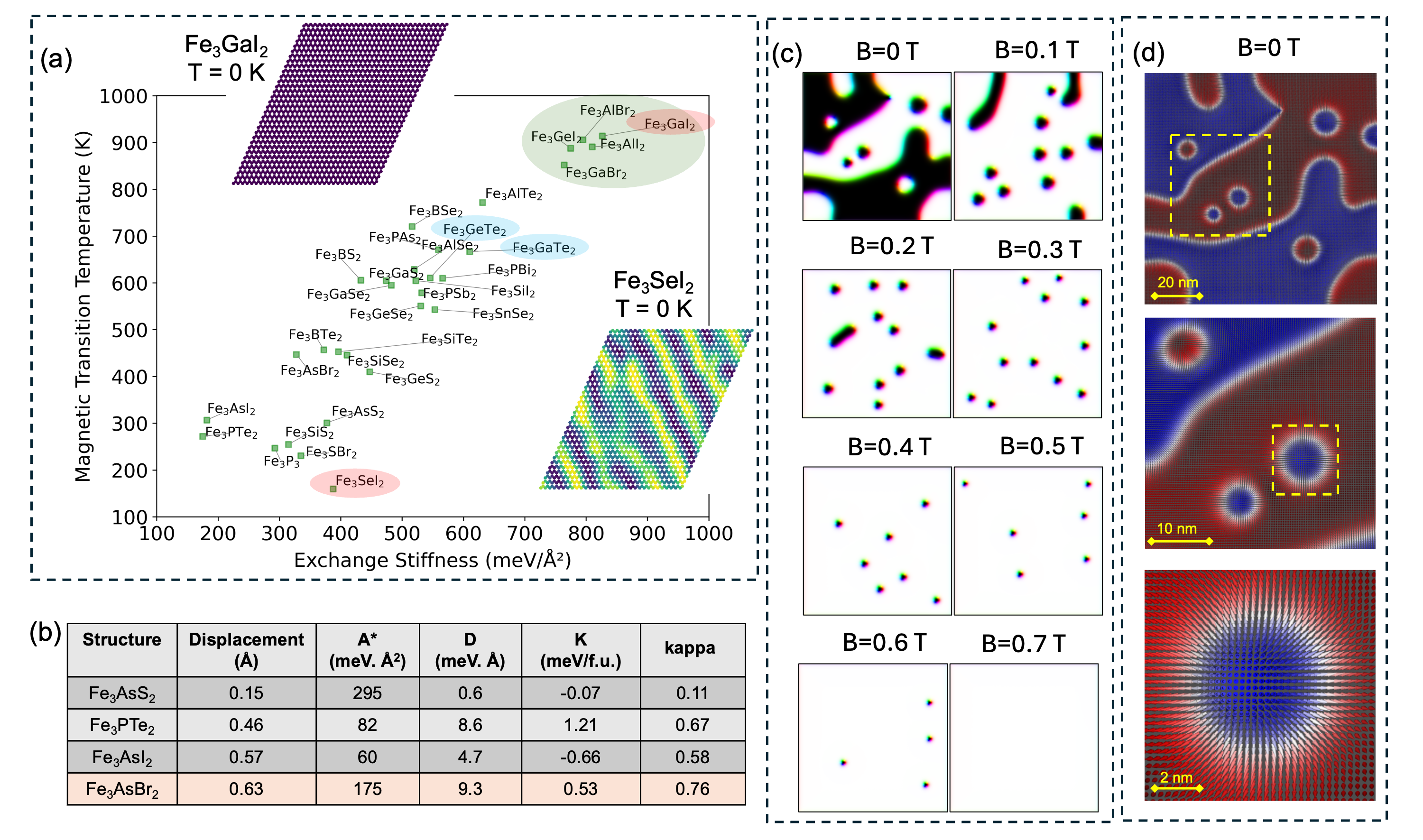}
  \caption{(a) Scatter plot of the magnetic transition temperature ($T_c$) vs. exchange stiffness ($A$), where the two synthesized compounds Fe$_3$GeTe$_2$ and Fe$_3$GaTe$_2$ are highlighted in blue. The green zone highlights halide compounds, which exhibit the highest magnetic transition temperatures, while the compounds with the highest ($T_c$ for Fe$_3$GaI$_2$) and lowest ($T_c$ for Fe$_3$SeI$_2$) values are highlighted in red. Two insets show the Monte Carlo spin texture at $T=0$ K for Fe$_3$GaI$_2$ and Fe$_3$SeI$_2$, where dark blue represents up-spin alignment and yellow represents down-spin alignment. Fe$_3$GaI$_2$ exhibits a collinear spin texture, whereas Fe$_3$SeI$_2$ shows spin spirals. (b) List of non-centrosymmetric compounds, ordered based on centrosymmetry-breaking displacements between Fe$_2$ and X, together with the corresponding micromagnetic quantities: exchange stiffness ($A$), spiralization constant ($D$), and magnetic anisotropy ($K$). The only combination that exhibits a topological spin texture is highlighted in orange. (c) Micromagnetic simulation for Fe$_3$AsBr$_2$ as a function of external magnetic field, where skyrmions emerge at $B=0$ T, shrink in size with increasing magnetic field, and vanish for fields larger than $B=0.7$ T. (d) Colored zero-field ($B=0$ T) spin texture of Fe$_3$AsBr$_2$, with the middle and bottom panels zoomed into the N\'eel-type skyrmionic texture. Red indicates spin-up alignment, while blue indicates spin-down alignment.
}
  \label{fig4}
\end{figure*} 

To gain insight into the origin of the observed exchange interactions and their peculiar correlations, we analyzed the orbital-resolved contributions to $J_{13}$ (Fig.~\ref{fig3}(f)), $J_{12}$ (Fig.~\ref{fig3}(g)), and $J_{22}$ (Fig.~\ref{fig3}(h)), where the dominant orbital channels are highlighted in orange. It is evident that the $d_{z^2}$–$d_{z^2}$ channel dominates the $J_{13}$ exchange interaction. Fig.~\ref{fig3}(i) schematically illustrates the spatial alignment of the $d_{z^2}$ orbitals on Fe$_{1}$ and Fe$_{3}$ atoms, which are vertically aligned, along with the adjacent $p_z$ orbital of the Ge atom. The double-headed orange arrows and associated values indicate the tight-binding hopping parameters between the shown orbitals. Due to this vertical alignment, a direct overlap between the $d_{z^2}$ orbitals of Fe$_{1}$ and Fe$_{3}$ is expected and confirmed by a strong hopping parameter of $t = 0.7$~eV. Additionally, the $p_z$ orbital of the Ge atom mediates an indirect interaction between the $d_{z^2}$ orbitals of Fe$_{1}$ and Fe$_{3}$, supported by a relatively strong hopping parameter of $t = 0.4$~eV. These findings suggest that $J_{13}$ is primarily governed by a direct exchange mechanism, while indirect exchange via the $p$-orbitals of the Ge atom also has a contribution. 

A similar analysis was performed for $J_{12}$, where the dominant exchange channel is found to be between $d_{z^2}$-$d_{xz}$ orbitals, as shown in Fig.~\ref{fig3}(j). In this case, a direct overlap between the $d$-orbitals of Fe atoms is spatially less probable, which is reflected in the very weak direct hopping parameter of $t = 0.03$~eV. However, the $p$-orbital of the neighboring Ge atom mediates an indirect exchange between the d-orbitals with a relatively strong hopping parameter of $t = 0.6$~eV. This indicates that $J_{12}$ is predominantly governed by an indirect exchange mechanism facilitated by $p$–$d$ hybridization. In the same manner, for the AFM $J_{22}$ interaction, the dominant exchange channel is also between $d_{z^2}$-$d_{z^2}$ orbitals, as shown in Fig.~\ref{fig3}(h). However, since the two Fe atoms involved in this interaction lie in the same $xy$-plane, a direct orbital overlap is not feasible. Therefore, the exchange interaction is expected to be mediated indirectly through the $p$-orbitals of neighboring Ge and Te atoms. 

These findings suggest that $J_{13}$ is less sensitive to structural parameters and uncorrelated with $J_{12}$ or $J_{22}$ due to its primarily direct exchange mechanism. In contrast, both $J_{12}$ and $J_{22}$ exhibit strong mutual correlation and dependence on structural features, consistent with their shared indirect exchange nature mediated by the $p$-orbitals of the $X$ and $Y$ atoms.

Figure~\ref{fig3}(d) presents the cutoff radius beyond which the strength of exchange interactions falls below 1~meV. For most compounds, this cutoff lies in the range of 7–8~\AA, indicative of long-range magnetic interactions typically observed in metallic systems. Interestingly, despite all compounds exhibiting metallic character (as confirmed by the electronic band structures in Fig.~S3), certain materials such as Fe$_3$PTe$_2$ and Fe$_3$SiS$_2$ show significantly shorter interaction ranges, with cutoff radii as low as 4.5~\AA.

Fig.~\ref{fig4}(a) presents a scatter plot of the magnetic transition temperatures ($T_c$), estimated via Monte Carlo simulations, versus the exchange stiffness ($A$). As expected, a general linear trend is observed, where higher exchange stiffness correlates with increased $T_c$. Notably, halide compounds (highlighted in green) exhibit higher $T_c$ values than those of the experimentally synthesized Fe$_3$GeTe$_2$ and Fe$_3$GaTe$_2$. This suggests that synthesizing Fe$_3$XY$_2$ compounds with halogens at the $Y$ site may lead to new 2D ferromagnets with record-high transition temperatures, enabling room-temperature spintronic applications. 

It is important to note that Monte Carlo simulations based on GGA-extracted exchange parameters tend to overestimate $T_c$ due to the known overestimation of exchange interactions within GGA. This limitation can be addressed by incorporating dynamical electronic correlations via DMFT, although DMFT is computationally prohibitive for high-throughput studies. Prior studies on the Fe$_n$GeTe$_2$ family show that the primary correction from DMFT is a renormalization of $J_{13}$ by approximately a factor of two, attributed to the correlation-induced broadening of Fe $d_{z^2}$ orbitals~\cite{ghosh2023unraveling}.

The two insets in Fig.~\ref{fig4}(a) depict the spin textures at $T = 0$~K for Fe$_3$GaI$_2$ (with the highest $T_c$) and Fe$_3$SeI$_2$ (with the lowest $T_c$) among the stable compounds. Fe$_3$GaI$_2$ exhibits a fully collinear ferromagnetic configuration, consistent with its strong ferromagnetic exchange interactions (e.g., $J_{13} = 110.6$~meV and $J_{12} = 43.5$~meV) and relatively weak antiferromagnetic interactions ($J_{22} = -3.3$~meV). In contrast, Fe$_3$SeI$_2$ exhibits non-collinear spin spirals at $T = 0$~K, arising from competing ferromagnetic and antiferromagnetic interactions. This frustration leads to a lower thermal stability of magnetic order and consequently a reduced $T_c$.

In Fig.~\ref{fig4}(b), systems with non-centrosymmetric structural ground states are shown, sorted by the vertical displacement between Fe$_2$ and $X$ atoms. The exchange stiffness ($A$) and the spiralization tensor ($\mathbf{D}$) were computed using the following expressions~\cite{borisov2024electronic}:

\begin{equation}
A = \frac{1}{2} \sum_{j \neq i} J_{ij} R_{ij}^2 \, e^{-\mu R_{ij}},
\label{A}
\end{equation}

\begin{equation}
D_{\alpha \beta} = \sum_{j \neq i} D_{ij}^{\alpha} R_{ij}^{\beta} \, e^{-\mu R_{ij}},
\label{D}
\end{equation}

where \( J_{ij} \) and \( D_{ij}^{\alpha} \) represent the isotropic and DMI exchange interactions between atomic sites \( i \) and \( j \), respectively, and \( R_{ij} \) denotes the distance between them. The exponential damping factor \( e^{-\mu R_{ij}} \) ensures convergence with respect to the real-space cutoff. To accurately determine \( A \) and \( \mathbf{D} \), the parameter \( \mu \) was varied between 2 and 4, and the results were extrapolated to \( \mu = 0 \) using a third-order polynomial fit~\cite{borisov2024tunable, borisov2024dzyaloshinskii}. In two-dimensional systems, the interfacial Dzyaloshinskii–Moriya interaction (DMI) typically dominates, leading to a characteristic form of the spiralization matrix. As a result, the spiralization tensor generally takes the form:

\[
\mathbf{D} =
\begin{bmatrix}
0 & D & 0 \\
-D & 0 & 0 \\
0 & 0 & 0
\end{bmatrix},
\]

where \( D \) denotes the scalar spiralization constant, expressed in units of meV\(\cdot\)Å. For each compound, the normalized exchange stiffness ($A^*$, where $J_{13}$ is halved), spiralization constant ($D$), and magnetic anisotropy energy ($K$) are listed, as key input parameters for micromagnetic simulations. To quantify the competition between these parameters, a dimensionless stability criterion $\kappa$ is commonly used to assess the likelihood of chiral magnetic states~\cite{rohart2013skyrmion}, defined as:
\begin{equation}
\kappa = \frac{\pi D}{4 \sqrt{A K}}.
\end{equation}
In general, $\kappa > 1$ (or close to 1) is considered favorable for stabilizing skyrmions. The calculated $\kappa$ values for each system are listed in the last column. Micromagnetic simulations showed trivial spin textures for all compounds except Fe$_3$AsBr$_2$, which is highlighted in orange.

Fig.~\ref{fig4}(c) illustrates the evolution of the spin texture in Fe$_3$AsBr$_2$ under varying external magnetic fields. Notably, skyrmion-like spin textures are present even at zero external field ($B = 0$~T). As the magnetic field increases, the skyrmions shrink and eventually vanish at fields above $B = 0.7$~T. Fig.~\ref{fig4}(d) presents a magnified view of the spin texture at $B = 0$~T, with two zoomed-in regions (highlighted by dashed yellow boxes) shown in the lower panels. The observed spin texture is consistent with Néel-type skyrmions, with a characteristic diameter of approximately 10~nm.

To confirm the topological nature of these spin textures, the winding number was computed using:
\begin{equation}
N_{\text{sk}} = \frac{1}{4\pi} \int \mathbf{m} \cdot 
\left( \frac{\partial \mathbf{m}}{\partial x} \times \frac{\partial \mathbf{m}}{\partial y} \right) 
\, dx \, dy.
\label{winding}
\end{equation}
The result yields a winding number close to \( N_{\text{sk}} = 1 \)~\cite{kim2020quantifying}, verifying the topologically non-trivial nature of the magnetic configuration and confirming the presence of Néel-type skyrmions.

\section{Methodology}
First-principles structural optimization, stability analysis, and calculation of magnetic ground state were conducted in the framework of density functional theory (DFT) \cite{kohn1965self,hohenberg1964inhomogeneous} using the Vienna \textit{Ab initio} Simulation Package (VASP) \cite{kresse1993ab,kresse1994ab,kresse1996efficiency,kresse1996efficient}. The exchange-correlation potential was treated using the generalized gradient approximation (GGA) functional in conjunction with the Perdew, Burke, and Ernzerhof (PBE) method \cite{perdew1996generalized}. The projector augmented wave method \cite{Blochl1994projector} was applied. A plane-wave basis set with a kinetic cutoff energy of 500 eV was used to expand the electronic wave function, and a vacuum space of at least 20 \AA~ was inserted along $c$-axis to prevent unrealistic interactions between periodic images. During structural optimization, the maximum force on each atom was less than  5$\times$10$^{-3}$ eV/\AA~. A~Gaussian smearing factor of 0.05 was taken into account. Brillouin zone (BZ) integration was performed by a $\Gamma$-centered $9\times9\times1$  uniform $k$-point grid for monolayers. To ensure the dynamical stability of the systems, the phonon dispersions were calculated, using the finite-displacement approach, implemented in the PHONOPY \cite{togo2015first} package. To analyze the thermal stability, \textit{ab initio}  molecular dynamics (AIMD) simulations were conducted, using a microcanonical ensemble (NVE), at constant temperatures of T = 500 K for a simulation period of 5 ps with 1 fs time steps. A $3\times3\times1$ supercell was used for phonon dispersions and AIMD. For the calculation of Heisenberg exchange interactions (Jij and Dij) and magnetic anisotropy energy (MAE), a $25\times25\times1$ $k$-point grid was utilized, and the spin–orbit coupling (SOC) was taken into account. All magnetic properties (Magnetic moments, Heisenberg exchange interaction, DMI, MAE, and atom resolved MAE) were calculated via the QuantumATK-Synopsys package version U-2022, employing an LCAO basis set, the "PseudoDojo" pseudopotential \cite{qatk1, qatk2}, a density mesh cutoff of 120 Hartree. The projection into localized Wannier Functions (WFs) was carried out using the Wannier90 package \cite{mostofi2008wannier90}, via VASP to Wannier90 interface. The WFs basis set comprised five d-orbitals of transition metal and three p-orbitals of X and Y atoms. The tight-binding hopping parameters were extracted from WFs. The TB2J \cite{he2021tb2j} package was utilized to calculate orbital resolved isotropic exchange interactions via Liechtenstein, Katsnelson, Antropov, and Gubanov (LKAG) \cite{liechtenstein1987local} formalism. The extracted exchange interactions and MAE were implemented in a Heisenberg Hamiltonian to calculate the magnetic ordering temperature by performing classical Monte Carlo (MC) simulations via UppASD code \cite{eriksson2017atomistic}. To achieve properly averaged properties, five ensembles within a supercell of $32\times32\times1$ were modeled, assuming periodic boundary conditions. Micromagnetic simulations were performed using MuMax3\cite{vansteenkiste2014,mulkers2017}. The unit cell dimensions were obtained from relaxed structures in \textit{ab initio} calculations and were used to form a grid of $256 \times 256 \times 1$ cells. The Gilbert damping parameter was set to $\alpha = 0.10$ as proposed in a previous study \cite{joos2023tutorial}, and uniaxial anisotropy was applied. The simulation time was 1 ns for each set of parameters, ensuring the system would reach an equilibrium state. The values for exchange stiffness ($A$), DMI ($D$), uniaxial anisotropy constant ($K_{u1}$), and magnetic moment per unit cell ($\mu$) were taken from \textit{ab initio} calculations and converted into the continuous material parameters used in micromagnetic simulations. The system was subjected to an external magnetic field applied perpendicular to the atomic plane, with field magnitudes ranging from 0~T to 1.0~T in steps of 100~mT.

\begin{acknowledgments}
B.S. acknowledges financial support from Swedish Research Council (grant no. 2022-04309 and grant No. 2018-07082). The computations were enabled by resources provided by the National Academic Infrastructure for Supercomputing in Sweden (NAISS) at UPPMAX (NAISS 2024/5-258) and at NSC and PDC (NAISS 2024/3-40) partially funded by the Swedish Research Council through grant agreement no. 2022-06725. B.S. and S. E. also acknowledge EuroHPC for awarding us access to EU2023D11-039 hosted by Karolina at IT4Innovations, Czech Republic. Artificial intelligence was used to improve the language and readability of this work.
\end{acknowledgments}

\section*{Data Availability Statement}
All data for stable compounds are presented in the main manuscript and supplementary materials. Data related to unstable compounds are also available from the corresponding author upon reasonable request.

\section*{Conflict of interest}
The authors declare that they have no conflicts of interest to disclose.

\nocite{*}
\bibliography{aipsamp}

\end{document}